\documentclass{article}

\usepackage[utf8]{inputenc}
\usepackage{url}
\usepackage{color}
\usepackage{caption}
\usepackage{subcaption}
\usepackage{graphicx}
\usepackage{multirow}
\usepackage{lineno}

\title{Gravitational Wave Detection and Information Extraction via Neural Networks}

\author{Gerson R. Santos$^{1}$, Marcela P. Figueiredo$^1$, Antonio de P\'adua Santos$^2$,\\ Pavlos Protopapas$^3$,Tiago A. E. Ferreira$^{*,1,3}$}

\date{%
    $^1$Departamento de Estat\'{i}stica e Inform\'{a}tica, Universidade Federal Rural de Pernambuco, Brasil\\%
    $^2$Departamento de F\'{i}sica, Universidade Federal Rual de Pernambuco, Brasil\\%
    $^3$Institute For Applied Computational Science, Harvard University, USA
}

\begin{document}

\maketitle




\begin{abstract}
Laser Interferometer Gravitational-Wave Observatory (LIGO) was the first laboratory to measure the gravitational waves. It was needed an exceptional experimental design to measure distance changes much less than a radius of a proton. In the same way, the data analyses to confirm and extract information is a tremendously hard task. Here, it is shown a computational procedure base on artificial neural networks to detect a gravitation wave event and extract the knowledge of its ring-down time from the LIGO data. With this proposal, it is possible to make a probabilistic thermometer for gravitational wave detection and obtain physical information about the astronomical body system that created the phenomenon. Here, the ring-down time is determined with a direct data measure, without the need to use numerical relativity techniques and high computational power.

\end{abstract}




\section{Introduction}

Einstein had predicted gravitational waves in 1916, but its experimental confirmation was done almost one century after. Among other tests, the detection of gravitational waves performs one of the main confirmations of the theory of general relativity \cite{misner1973gravitation,hartle2003gravity, carroll2019spacetime}. Such measurements of gravitational wave signals can reveal crucial information about the nature of spacetime.  Because it is a very tenuous signal, the detection of gravitational waves requires complex experimental apparatus. In this context, significant advances in the discovery of gravitational waves were made in the last 50 years. Those developments have brought new technologies in data analysis, Optics and Quantum Mechanics, Laser experiments,  materials science, cryogen, and other areas \cite{tech, tech2}. Possible sources of gravitational waves have been sought for decades, the most likely binary stars, pulsars, instability in type II supernovae. In 1975, Hulse and Taylor discovered a binary system formed by a neutron star and a pulsar orbiting a common center (PSR 1913+16) \cite{hulse1975discovery, hulse1994discovery,taylor1994binary}. Studies conducted by Hulse and Taylor demonstrate that this object can be considered strong evidence of the emission of gravitational waves. The discovery of binary systems formed by compact stellar objects has opened up vast possibilities for gravitational wave detection.

Recently, the Laser Interferometer Gravitational-Wave Observatory (LIGO) performed the first measure of gravitational waves, GW150914 \cite{abbott2016observation}.  The LIGO used principles of interferometry based on the Michelson-Morley apparatus.  So far, LIGO Scientific Collaboration has performed three observing runs in which several gravitational wave emission events have been detected. Data provided by LIGO show that gravitational waves were produced by a merge of binary systems formed by compact stellar objects, such as black holes and neutron stars.

The merge of compact stellar binary systems such as those formed by black holes or neutron stars occurs in three phases. At first, the system presents a common center describing circular orbits. This phase is known as the inspiral phase. At this stage, the system loses orbital energy in the form of gravitational waves. In a second phase, black holes come together, forming a single black hole. In the third phase, named ring-down, the merged black holes take on the equilibrium state as a spinning black hole known as the Kerr black hole. A complete description of the inspiral, merged, and ring-down stages can be seen in computational simulations of each phase \cite{abbott2016observation}. In general, the computational simulations require sophisticated Numerical Relativity techniques with high computational cost and complex algorithms. On the other hand, the signal measured by LIGO requires a detailed inspection of the data using in-depth techniques of data analysis and computational methods.

%

 This article has proposed an application of the artificial neural networks to detect the gravitational wave event in the LIGO data and to obtain the physical parameters of this system. The objective is to complement the data analyses carried out by the LIGO scientific consortium. Besides, we show that simple artificial neural networks, along with statistical data analysis, are useful tools complementary to the study of big data of gravitational waves produced by compact stellar systems such as black holes. In the next section, we describe a simple model to understand the gravitational waves. After that, Section \ref{sec:ANN} shows the Artificial Neural Network approach used here. Section \ref{sec:Data_Exp} presents a discussion about the data and the experimental results with the measurement of the ring-downs times. Finally, in Section \ref{sec:conclusion} are shown the conclusions of the article.

\section{A Simple Phenomenological Model}\label{sec:model}

\begin{figure}
    \centering
    \includegraphics[width = \textwidth]{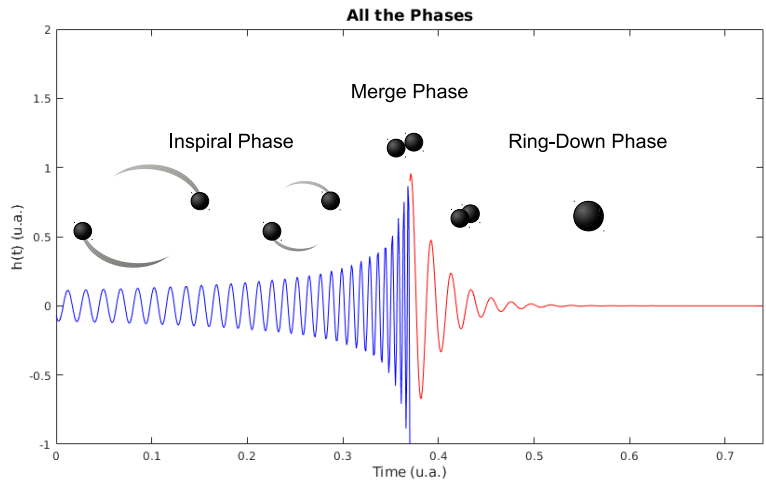}
    \caption{All the phases for a binary system. Gravitational wave profile from Equation (\ref{strain}), the inspiral phase --- blue line. Ring-down stage profile from Equation\ref{damping1}, the ring-down phase --- red line.}
    \label{fig:GW_phases}
\end{figure}

From a phenomenological point of view, we assume a naive model, but a detailed description of gravitational waves and their detection can be found in \cite{maggiore1, maggiore2}. The gravitational wave produces a strain in the interferometer \cite{mathur2017analysis,rubbo2007hands,ligo2017basic}. At the inspiral stage, the strain $h(t)$ produced in the interferometer is correlated with the gravitational waveform.  For this model the strain behavior can be described by \cite{mathur2017analysis}
\begin{equation}
    h(t) = \mathcal{A}(t)\textsl{cos}(\Phi(t)),
    \label{strain}
\end{equation}
where $h(t)$ is the gravitational waveform (or strain), $\mathcal{A}(t)$ is the amplitude and $\Phi(t)$ is the gravitational wave phase. The time-dependent amplitude is described as following
\begin{equation}
    A(t) = \frac{2(G\mathcal{M})^{5/3}}{c^4r}\Bigg(\frac{\pi}{P_{gw}(t)}\Bigg)^{2/3},
    \label{amp1}
\end{equation}
where $G$ is the Newtonian gravitation constant, $c$ is the speed of light, $r$ is the luminosity distance to the binary and   $\mathcal{M}$ is the chirp mass given by
\begin{equation}
    \mathcal{M} = \frac{(M_1M_2)^{3/5}}{(M_1+M_2)^{1/5}}.
\end{equation}
$M_1$ and $M_2$ are the black holes masses. The gravitational wave period is related to the period of the binary system by
$P_{gw}(t) = \frac{P_{orb}(t)}{2}$
\begin{equation}
    P_{orb}(t) = \Bigg(P_o^{8/3} -\frac{8}{3}kt\Bigg)^{3/8},
    \label{Porb}
\end{equation}
where the constant $k$ is given as following
\begin{equation}
    k =\frac{96}{5}(2\pi)^{8/3}\Bigg(\frac{G\mathcal{M}}{c^3}\Bigg)
    \label{k}
\end{equation}
From Equation (\ref{Porb}) and Equation (\ref{k}) we can determine the orbital period $P_{gw}(t)$. The gravitational wave phase is given by the integration of the orbital period as
\begin{equation}
    \Phi(t)= \Phi_0 + 2\pi\int^t_0 \frac{1}{P_{gw}(t')}dt'
    \label{phase}
\end{equation}
The gravitational wave profile is determined using Equation (\ref{amp1}) and Equation (\ref{phase}) in Equation (\ref{strain}). This behavior for inspiral stage can be seen in Figure \ref{fig:GW_phases} (blue line).


Now we consider the ring-down stage. After the merge of black holes, a final black hole with spin is formed. The gravitational waves after the merge fade out. Similar to the previous case, we use a phenomenological model to describe the behavior of the gravitational wave. At this stage, gravitational wave oscillations can be seen as a damped harmonic oscillator as follows \cite{marion2013classical},

\begin{equation}
    h(t) = B e^{\frac{-t}{\tau}}\textsl{sin}(\omega t +\delta_0) + D
    \label{damping1}
\end{equation}
where $B$ and $D$ are constants, $\delta_0$ is an initial phase and $\tau$ is the damping constant. The profile of the function can be seen in the Figure \ref{fig:GW_phases} (red line).


\section{Artificial Neural Network Classify}\label{sec:ANN}

An Artificial Neural Network (ANN) is a computational model biologically inspired. It tries to simulate the information process of a brain ~\cite{franchini2018artificial,yalccinreconfigurable}. There are many types of ANN, but in general, an ANN mathematically is an universal function approximator~\cite{Cybenko1989}.

The most popular ANN model is the Multi-Layer Perceptron (MLP). The MLP is a network described in layers, where the first layer is the input of the network, the last layer is the output, and the intermediate layers are the hidden layers. When the number of hidden layers is vast, the MLP is naturally into the Deep Learning branch~\cite{Goodfellow2016,Kamath2019}.

In general, problems with a high amount of observations and without analytic model known are excellent candidates to systems based on ANN solutions. The ANN's features of learning and information generalization are the base to generate solutions for a large class of problems, like classification, recognition,  clustering, forecasting, regression, data mining, among others~\cite{Abiodun2018}.

In particular, for the classification problem, the computational methods attempt to classify new objects according to some similarity measured from an existing group of classes. In this way, an ANN of MLP type will need to learn the similarity characterizes of the known object to classify new objects. The MLP learns these characterizes through a supervised learning process. In this learning process, the ANN receives labeled examples of all classes,  where the ANN can learn and generalize the similarity characterizes of the classes. For binary problems, where there are only two classes, an ANN  would only need one output. This output would be $0$ if the input were of the class $\mathcal{X}$ and $1$ if the input were of the class $\mathcal{Y}$, for example.

\begin{figure}
    \centering
    \includegraphics[scale=0.65]{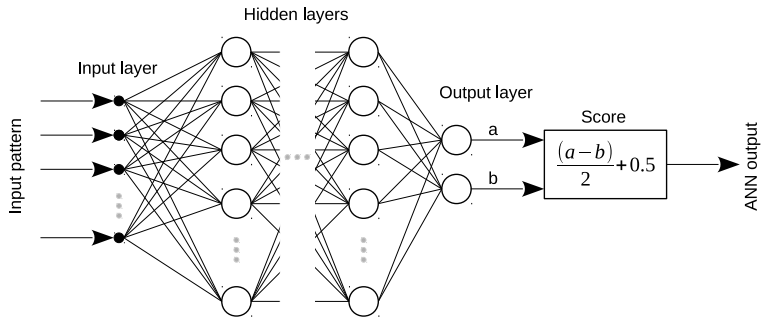}
    \caption{MLP scheme for binary problems with two outputs, $a$ and $b$, and the score computation.}
    \label{fig:ANN2outputs}
\end{figure}

However, it is possible to enhance the ANN discrimination by putting two outputs in the last ANN layer, the outputs $a$ and $b$, as schemed in Figure \ref{fig:ANN2outputs}. Now, for example, if the input were of the class $\mathcal{X}$, then the desired ANN outputs were $a=0$ and $b=1$, and if the input were of the class $\mathcal{Y}$, then the desired ANN outputs were $a=1$ and $b=0$.  With these two outputs, it is possible to define a score as,
\begin{equation}\label{eqn:score}
score = \frac{a-b}{2}+0.5
\end{equation}

\textbf{}

In this way, the desired ANN output for class $\mathcal{X}$ will have a score that will tend to 0 ($score \rightarrow 0$), and the desired ANN output for class $\mathcal{Y}$ will have a score that will tend to 1 ($score \rightarrow 1$). Thus, it will be possible to build a score distribution for each class, and the discrimination capability of the ANN can be measured with the use of the Kolmogorov-Smirnov (KS) distance~\cite{Frederick2006}. This KS distance is the maximum distance between two empirical cumulative distribution functions (ECDF), that when normalized, is the base for the statistical Kolmogorov-Smirnov test or KS test~\cite{Frederick2006}. The KS test is used to decide if a sample comes from a population with a specific distribution. Therefore, the KS distance or KS test also can be used to verify if two empirical distributions are statistically equal or different. The KS distance will be $1$ for the perfect separability. 


%

The ANN training phase generates these two score distributions. After that, when the ANN receives news data to analyze, a score value will be computed. There are two forms to analyze this new score value. In the first methodology, in the training phase, a threshold score value is determined. Score values less than this threshold will be classified as class $\mathcal{X}$ and values greater than this threshold will be classified as class $\mathcal{Y}$. The second methodology is to consider the empirical score distribution like a pertinence function, where it is possible to define two fuzzy sets, and a fuzzy classification system can be employed to classify the new score data. Thus, the score value can be viewed like a thermometer, where low score values imply the class $\mathcal{X}$, and high score values imply class $\mathcal{Y}$. The idea of this thermometer is employed here.

\section{The Data and Experimental Results}\label{sec:Data_Exp}


Here, we will be examining the nine first Black-Hole Black-Hole (BBH) events observed by LIGO. These nine BBH merger gravitational waves were measured by two laboratories, one in Livingston (Louisiana) and another in Hanford (Washington), both labs in the USA. We collected all data of BBH merger gravitational waves from the LIGO-Virgo Gravitational Wave Open Science Center (GWOSC)\footnote{https://www.gw-openscience.org/about/}  at $4096$ Hz. All data were processed with the computational tool PyCBC~\cite{Nitz2018} in the same way that explained by the LIGO-VIRGO GWOSC tutorial~\footnote{https://www.gw-openscience.org/tutorials/}. However, only eight events of the Hanford data at a time were used to train the proposed model based on ANN. The one Hanford event not used to train the ANN, and all Livingston events were used to testing the ANN classifier.

With the Handford GW data processed, a wave signal was fitted in the region of the data where the LIGO discovered gravitational wave events. Eight fitted waves were used to create the positive class examples (gravitational wave event) of the ANN training set. For each one of these eight fitted waves, were created 5,000 positive patterns with a window of 0.1s of length, or 410 points, randomly located on the fitted waves. As the negative patterns, a window of the same time length (or 410 points)  was randomly situated on the processed GW data in places where only exist noise.  Here, it was also generated 5,000 patterns of negative class (no gravitational wave event).

\begin{figure}
    \centering
    \includegraphics[width=0.8\textwidth]{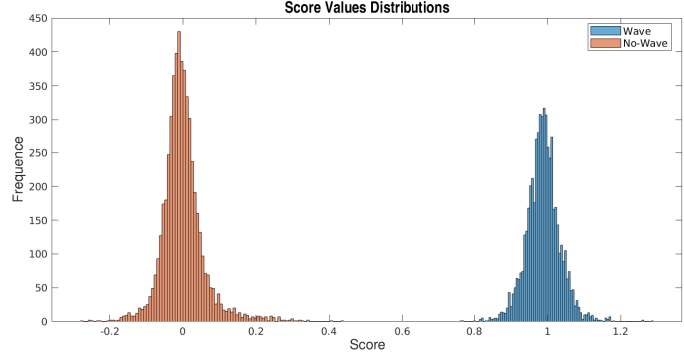}
    \caption{A typical score values distribution for trained ANN classify.}
    \label{fig:distroScoresANN}
\end{figure}

With a data set of 10,000 patterns (5,000 patterns of each class, positive and negative event), a committee of 500 MLP artificial neural networks was trained.  Each ANN of this committee has the same architecture with 410 inputs, one hidden layer with ten neurons with a hyperbolic tangent activation function, and two linear outputs. For each ANN, a score value (Equation (\ref{eqn:score})) is calculated, and a mean score can be computed. For the negative class, the $score \rightarrow 0$ , and for the positive class, the $score \rightarrow 1$. For all experiments, the ANN committee generated two distribution (one for the class ``wave'' and another for the class ``no-wave''), where it was measured a value of distance $KS = 1$. This $KS$ value indicates a perfect separability between the two class analyzed. Figure~\ref{fig:distroScoresANN} shows a typical score distribution generated by the ANN committee. 

\begin{figure}[t!]
     \begin{subfigure}{0.5\textwidth}
         \centering
         \includegraphics[width=1\textwidth]{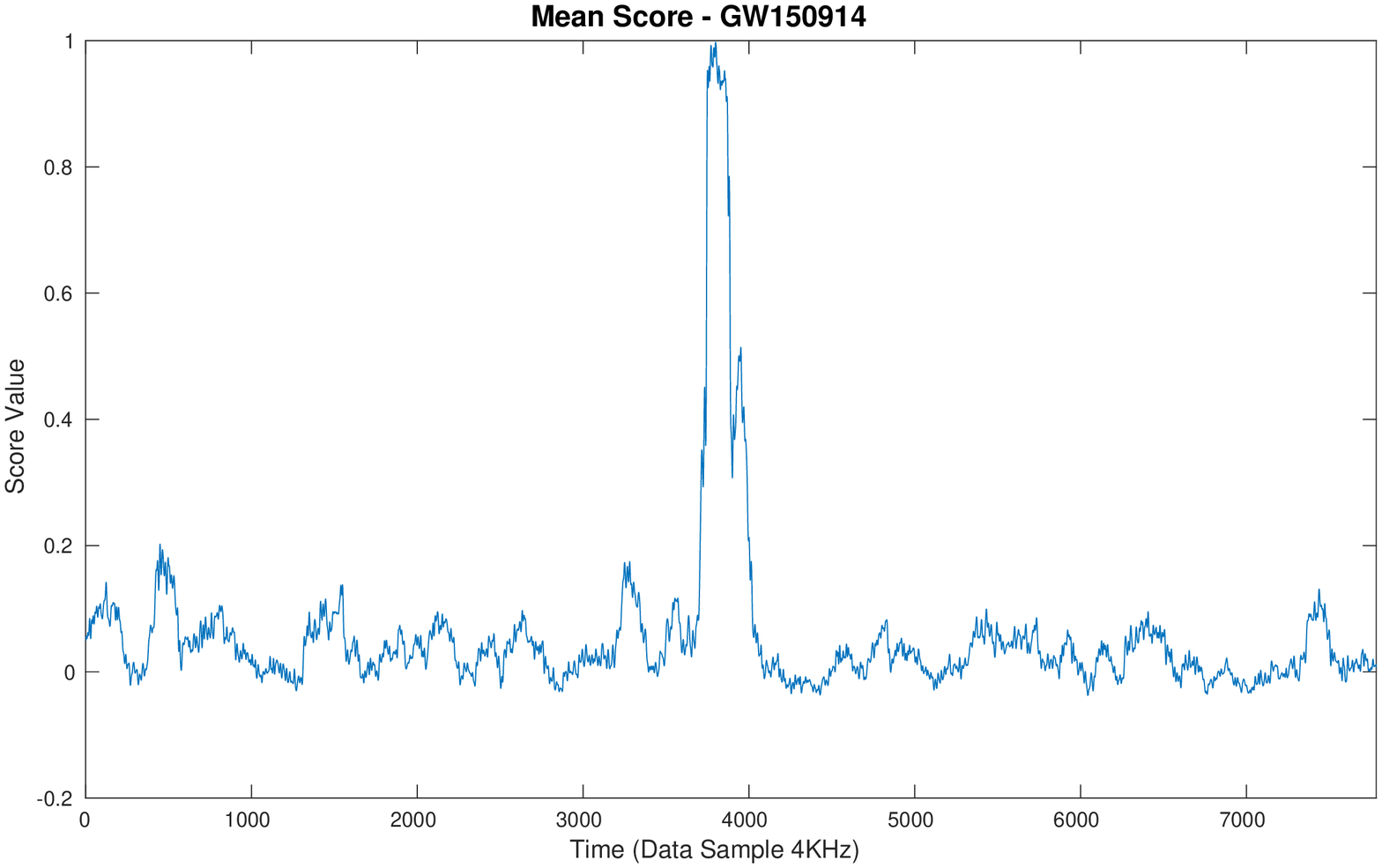}
         \caption{Handford.}
         \label{fig:resultScore_a}
     \end{subfigure}
     ~ 
     \begin{subfigure}{0.5\textwidth}
         \centering
         \includegraphics[width=1\textwidth]{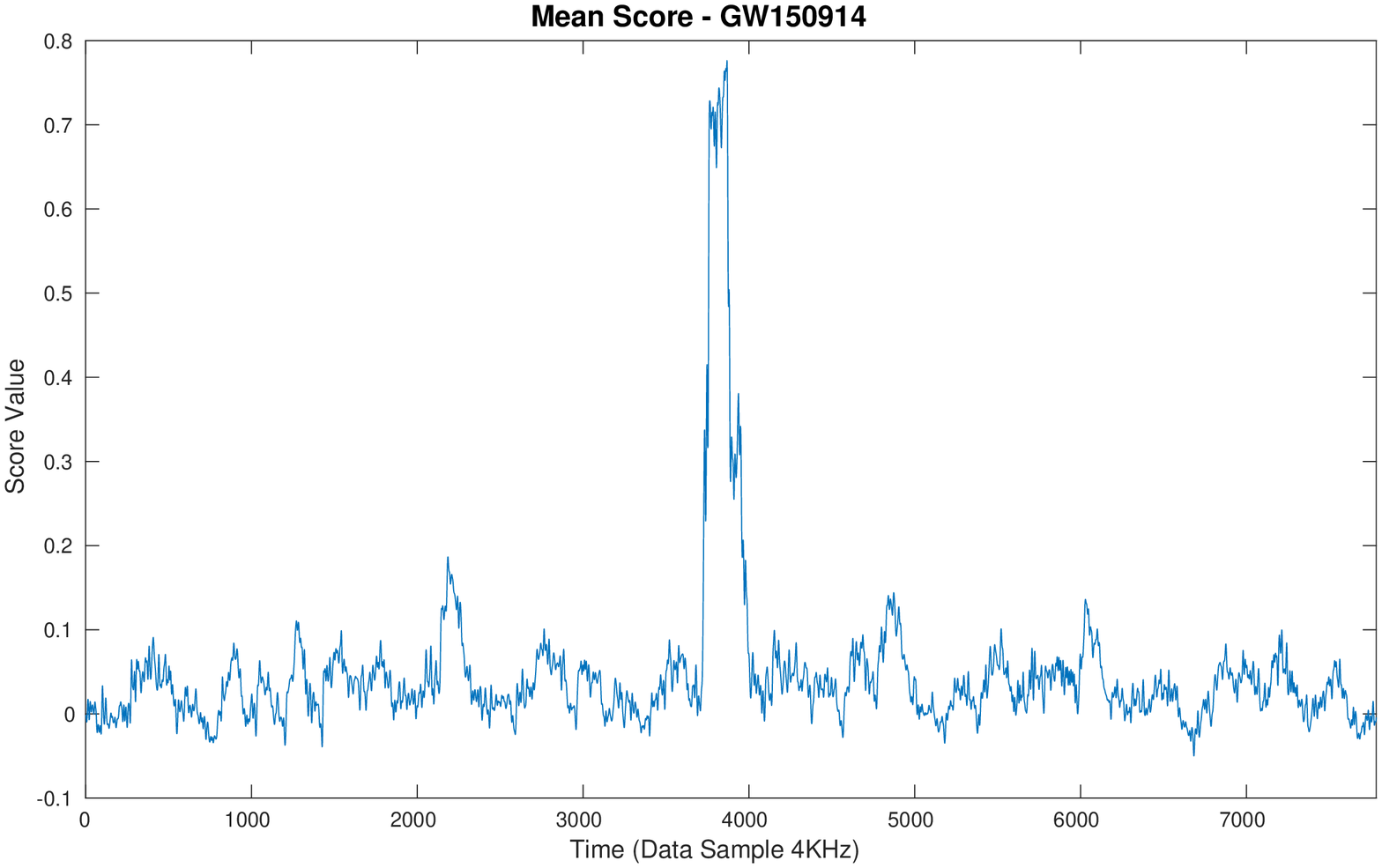}
         \caption{Livingston.}
         \label{fig:resultScore_b}
     \end{subfigure}
     ~
     \begin{subfigure}{0.5\textwidth}
         \centering
         \includegraphics[width=1\textwidth]{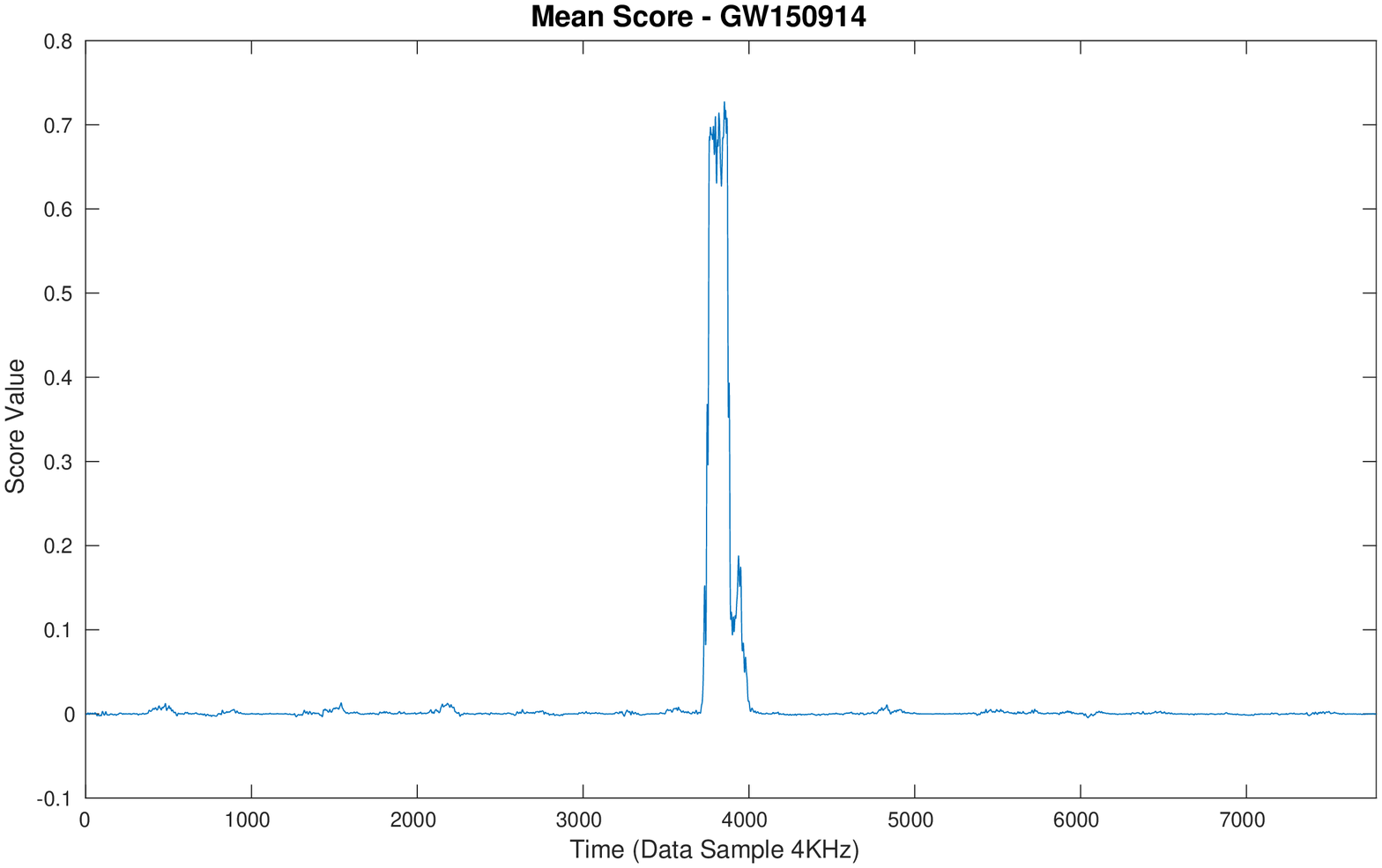}
         \caption{Coincidence.}
         \label{fig:resultScore_c}
     \end{subfigure}
     \begin{subfigure}{0.5\textwidth}
         \centering
         \includegraphics[width=1\textwidth]{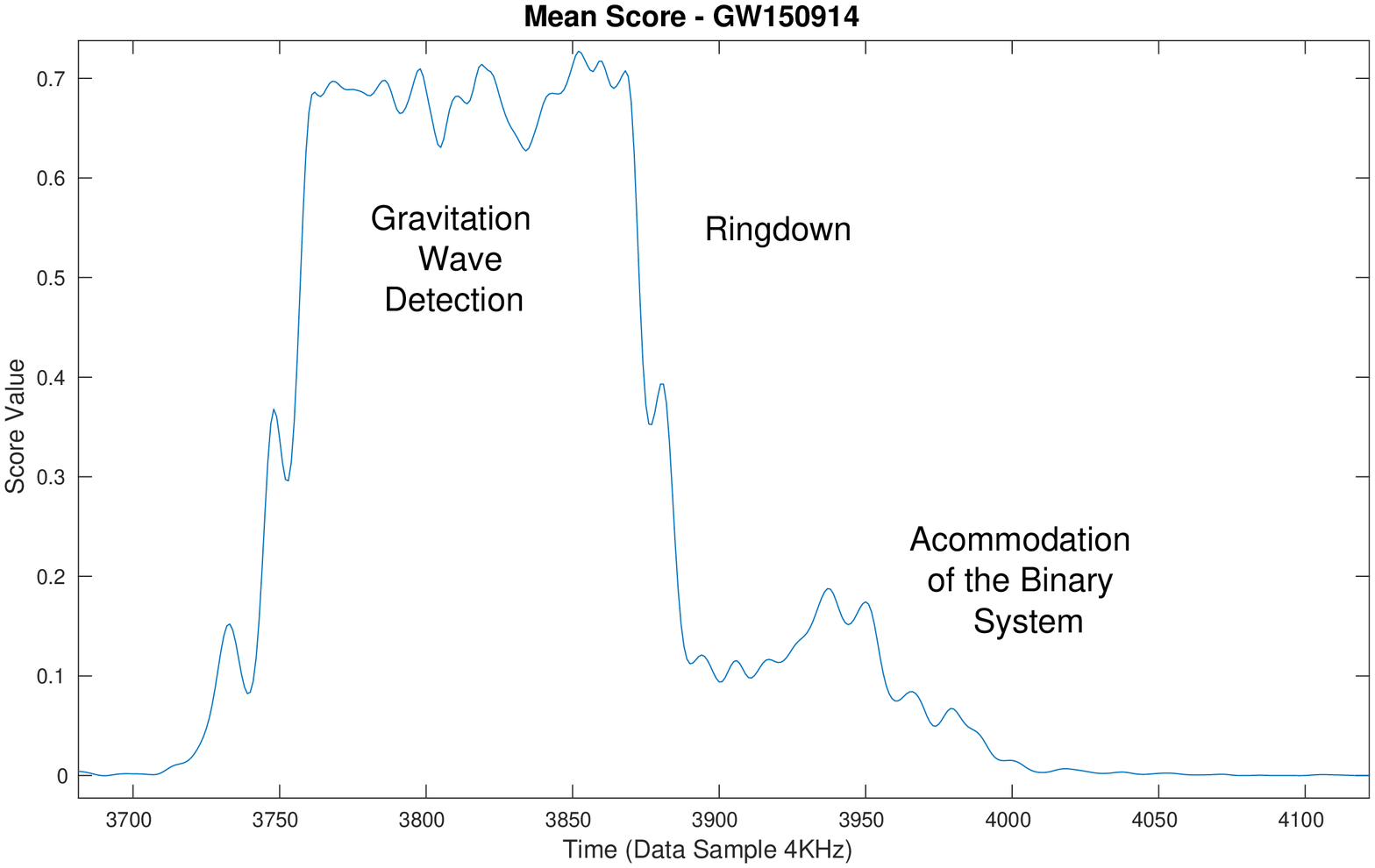}
         \caption{Coincidence Zoom.}
         \label{fig:resultScoreZoom}
     \end{subfigure}
        \caption{The time evolution of the coincident score for the GW150914 event from Handford and Livingston data.}
        \label{fig:resultScore}
\end{figure}

After the training process, the ANN committee is tested with the GW signal not used in training (both labs of Handford and Livingston). With a sliding window of 410 points (or $\approx 0.1$ s), a time-ordered sequence of input patterns is created and applied to the ANN committee. For all tests, the committee was able to determine when the GW event started and finished.  A typical output of the committee with the mean score value is presented in Figure~\ref{fig:resultScore} for the LIGO event GW150914 (Figure~\ref{fig:resultScore_a} is Handford and Figure~\ref{fig:resultScore_b} is Livingston). How the GW event is the same for both labs, it is possible to make a coincidence signal, Figure~\ref{fig:resultScore_c}, where a significant parcel of the noise is filtrated.

Magnifying Figure \ref{fig:resultScore_c} in the region where the wave occurred, Figure~\ref{fig:resultScoreZoom}, it is possible to observe the three regions:
\begin{enumerate}
\item A region whit high score value, indicating the gravitational wave detection in the spiral phase;
\item A region whit score value no zero, but lesser than the maximum score value. This region is probably the accommodation of the black hole binary system after the collision. A reverberation transient;
\item An intermediate region between the other two regions. This region is the beginning of the ring-down phase.
\end{enumerate}

\begin{figure}
    \centering
    \includegraphics[width=0.8\textwidth]{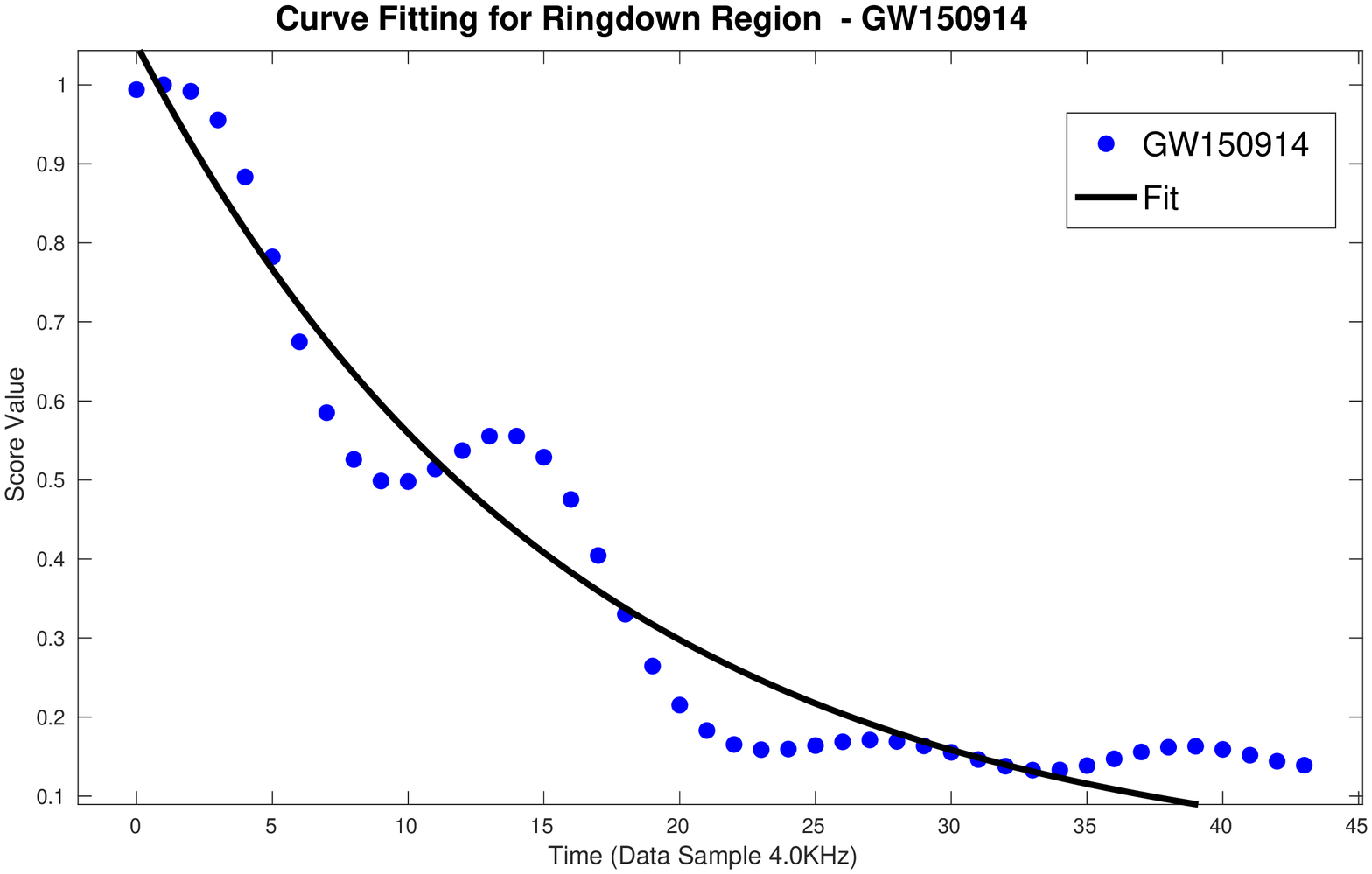}
    \caption{Curve fitting of the score values by $h(t) \propto \exp(\frac{-t}{\tau})$ for the ring-down region of the GW150914 event.}
    \label{fig:ringdownAdjust}
\end{figure}

Therefore, as presented by Equation~(\ref{damping1}), $h(t) \propto \exp(\frac{-t}{\tau})$, the decay of the GW amplitude in the ring-down stage, where $t$ is the time, and $\tau$ is the ring-down time. Figure~\ref{fig:ringdownAdjust} shows a exponential decay fitting for the GW150914 score data, in the region of ring-down indicated in Figure \ref{fig:resultScoreZoom}.  In the same way, it is possible to measure the ring-down time for every GW event observed from the score values. The Table~\ref{tab:ringdownTimes} presents the ring-down times measured for the nine GW observation by LIGO. The description of the parameters of binary systems formed by compact stellar objects that produced gravitational waves can be found in \cite{GW_data} and the most significant gravitational wave events from black holes merger can be found in \cite{abbott2016observation, abbott2016ligo, ligo2017cientific, abbott851and, scientific119virgo}.

\begin{table}
\small
\centering
\caption{Calculated ring-down times for the LIGO Black Hole -- Black Hole gravitational waves events.}\label{tab:ringdownTimes}
\medskip
\begin{tabular}{cccc}
\hline
\multirow{2}{*}{GW Event} & \multicolumn{2}{c}{Ring-down Times} & \multirow{2}{*}{LIGO Ring-down Times (ms)}\\
\cline{2-3}
& Times (ms) & 95\% Confidence bounds  & \\
\hline
GW150914 & 3.87	& (3.50; 4.24) & $\sim 4$ \\
GW151012 & 3.80 & (3.42; 4.20) & --- \\
GW151226 & 1.58 & (1.30; 1.86) &	$\sim 1.3$ \\
GW170104 & 2.76 & (2.36; 3.18) &	2.5 to 3.2 \\
GW170608 & 1.37 & (1.07; 1.67) &	1.0 to 1.4 \\
GW170729 & 3.07 & (2.56; 3.58) &  ---\\
GW170809 & 1.61 & (1.46; 1.76) &  ---\\
GW170814 & 3.33 & (2.78; 3.88) & 3.1 to 3.6 \\
GW170823 & 3.51 & (3.24; 3.77) & ---\\
\hline
\end{tabular}
\end{table}



\section{Conclusions}\label{sec:conclusion}

This article proposed a simple approach based on Artificial Neural Networks to recognize a gravitational wave event from the LIGO data, and from this recognition time solved, to extract the ring-down time information of the binary Black Hole system. With the proposal methodology, it was possible to define the ring-down time for all BBH events, including the events where the LIGO cannot calculate its ring-down time. Computationally, this proposal does not need high-performance computing. This proposal has a low computational cost, where a simple laptop or desktop can execute all analysis.

The experiments have shown that the MLP ANN committee was able to classify with high accuracy the presence or absence of the gravitational waves in the LIGO data. The KS distance equals to 1 reached by the proposed classifier demonstrates a perfect separability between the two classes (wave and no wave). In this way, allied with the $score$ definition (Equation~(\ref{eqn:score})), when the ANN reaches high values of the score, this implies in high probability to exist a gravitational wave event. 

All experiments were employed for both LIGO lab data, the lab of Handford, and the lab of Livingston.  How the gravitational wave detected by both labs are the same event, it is possible to make a coincidence analysis of the scores. It was possible to enhance the ration signal noise with is coincidence analysis, where for all gravitational wave events analyzed, the same typical behavior was observed. This behavior was an initial gravitational wave from the spiral phase of the binary system, a collision region of the two black-holes with decaying of the score value, and a small increase of the score signal after the decaying. This last slight increase of the score values comes probability from accommodation or reverberation of the resultant black-hole formed by the union of the two original black-holes. 

Therefore, the proposal of this article is a shallow computational cost form to analyze the LIGO data, where here it was searched for the gravitational wave events. However, in a more extensive view, maybe it also is possible to search for other kinds of signals in the LIGO,  like an earthquake signal or other geological events, and extract information.

\section*{Acknowledge}
To the Science and Technology Support Foundation of Pernambuco (FACEPE) Brazil, Brazilian National Council for Scientific and Technological Development (CNPq) and Coordena\c{c}\~{a}o de Aperfei\c{c}oamento de Pessoal de N\'{i}vel Superior - Brasil (CAPES) - Finance Code 001 by financial support for the development of this research.

\bibliographystyle{naturemag}
\bibliography{article}





\end{document}